\documentclass{article}
\usepackage[a4paper, left=1.5cm, right=1.5cm, top=2.5cm, bottom=2.5cm]{geometry}
\usepackage{graphicx} 
\usepackage{amsmath}
\usepackage{amssymb}
\usepackage{xcolor}
\usepackage{authblk} 
\usepackage{cite}

\title{Gaussian Planck Relics are Ruled-Out as Dark Matter by LIGO}

\author[1]{Oem Trivedi\thanks{oem.trivedi@vanderbilt.edu}}
\author[2]{Abraham Loeb\thanks{aloeb@cfa.harvard.edu}}

\affil[1]{Department of Physics and Astronomy, Vanderbilt University, Nashville, TN 37235, USA}
\affil[2]{Astronomy Department, Harvard University, 60 Garden St., Cambridge, MA 02138, USA}

\date{\today}

\begin{document}

\maketitle

\begin{abstract}
Quantum Gravity models suggest that singularities in gravitational collapse can be replaced by bounces, leading to the formation of Planck star remnants (PSR) that survive as stable relics once the black holes evaporate to the Planck mass. Recently, we proposed that such relics can be a viable candidate for dark matter. Here we show that LIGO's upper limit on the gravitational wave background rules out a formation pathway of Planck mass relics as dark matter from Gaussian initial conditions. This leaves non-Gaussian primordial fluctuations as the only viable channel for making Planck mass relics as dark matter.
\end{abstract}

\section{Introduction}

Few problems captivate as much attention in gravitational physics as the end state of matter undergoing complete collapse. Within the framework of classical general relativity, such a collapse inevitably culminates in the formation of a spacetime singularity \cite{sb1hawking1972black,sb2bekenstein1973black,sb3bekenstein1974generalized,sb4tipler1977singularities,sb5krolak1978singularities,sb6clarke1985conditions,sb7penrose1969gravitational}. At densities approaching the Planck scale, $\rho_{\rm Pl} \sim (c^5/h G^2) \sim 10^{93}\,{\rm g\,cm^{-3}}$, it is widely expected that classical dynamics must give way to quantum gravitational effects, which may prevent the singular outcome and instead trigger novel phenomena \cite{sb8joshi2002cosmic,sb9senovilla20151965,sb10penrose1965gravitational,sb11penrose1999question,sb12hawking1970singularities,sb13Ong:2020xwv,sb14Vagnozzi:2022moj,sb15joshi2011recent,sb16janis1968reality,sb17Joshi:2011zm}. Among the various candidate frameworks for quantum gravity, Loop Quantum Gravity (LQG) \cite{lqg3fRovelli:1994ge,lqg1Ashtekar:2004eh,lqg2Garay:1994en,lqg4Ashtekar:2011ni,lqg5Bern:2010ue,lqg6Rovelli:1997yv} has provided a concrete setting in which curvature singularities can be replaced by non-singular "quantum bounces". This has been demonstrated most explicitly in Loop Quantum Cosmology (LQC) \cite{lqc1Ashtekar:2006rx,lqc2Agullo:2016tjh,lqc3Bojowald:2005epg,lqc4Ashtekar:2011ni,lqc5Agullo:2023rqq,lqc6Ashtekar:2008zu,lqc7Banerjee:2011qu,lqc8Wilson-Ewing:2016yan,Nojiri:2005sx}, where both the Big Bang and black hole singularities are naturally avoided.
\\
\\
When collapse drives matter to the Planck density, quantum geometry corrections in LQG may produce a bounce while leaving the event horizon intact such that the bounce remains invisible to external observers. The natural implication is the existence of stable, long-lived remnants rather than divergent singularities. These objects often referred to as Planck stars, were introduced in the LQG context as finite sized cores whose radius approaches the Schwarzschild radius when the black hole mass reaches the Planck scale, $M_{\rm Pl} \sim \sqrt{h c/G} \sim 10^{-5}\,{\rm g}$ \cite{pl1Rovelli:2014cta,pl2Rovelli:2024sjl,pl3Rovelli:2017zoa,pl4Scardigli:2022jtt,pl5Wilson-Ewing:2024uad,pl6Barrau:2014hda,pl7Tarrant:2019tgv}. Externally, such remnants are virtually indistinguishable from evaporated black holes yet they possess a non-singular quantum gravitational interior.
\\
\\
This perspective has important implications for cosmology and dark matter. Primordial black holes (PBHs), seeded by large-amplitude curvature fluctuations in the early universe \cite{pbh1pbhzel1966hypothesis,pbh2hawking1971gravitationally,pbh3carr1974black,pbh4carr1975primordial,pbh5chapline1975cosmological,pbh6hawking1975particle,pbh7hawking1974black,pbh8khlopov1980primordial,pbh9polnarev1985cosmology,pbh10khlopov2010primordial,pbh11carr2016primordial,pbh12carr2020primordial,pbh13carr2024observational}, would undergo Hawking evaporation over cosmic timescales. However, if evaporation halts at the Planck scale due to quantum backreaction, these PBHs would not vanish completely but would instead leave behind relics of Planck mass. Such relics are natural candidates for cold dark matter as they are compact, stable, non-radiating, and interact only gravitationally, properties that align with all astrophysical and cosmological requirements for dark matter.
\\
\\
In our recent work \cite{psr1}, we investigated this possibility in detail, showing that Planck star remnants can arise generically within the LQC framework and may be produced abundantly enough to explain the observed dark matter density. The present paper continues this line of inquiry by considering the observational constraints on this scenario. In particular, we examine the collapse fraction of PBHs required to yield the correct relic abundance, the induced stochastic gravitational wave background associated with their formation and the crucial role of primordial non-Gaussianities. Section 2 provides a brief overview of the Planck relic dark matter idea, Section 3 contains the detailed analysis of observational bounds, and we conclude with a discussion in Section 4.
\\
\\
\section{Planck Mass Relics as Dark Matter}
The concept of Planck star remnants arises naturally when quantum gravitational corrections are applied to the problem of complete gravitational collapse. Within the effective dynamics of LQC, the evolution of the scale factor $a(t)$ for a homogeneous interior is governed by the modified Friedmann equation
\begin{equation}
H^{2} = \left(\frac{\dot{a}}{a}\right)^{2} = \frac{8\pi G}{3} \rho \left(1 - \frac{\rho}{\rho_{c}}\right)
\end{equation}
where $\rho$ is the matter density and $\rho_{c}$ is the critical density determined by the underlying loop quantum geometry, typically of order the Planck density. While the standard Friedmann equation allows for indefinite growth of $\rho$ as $a \to 0$, the correction term enforces a bounce at $\rho = \rho_{c}$, ensuring that the scale factor never vanishes. In the context of black hole interiors, this mechanism prevents the formation of singularities and instead produces a finite-density core.
\\
\\
If such collapse occurs within a black hole horizon, the bounce remains hidden from external observers due to the causal structure of spacetime. The result is a long-lived, non-singular object of roughly Planck mass and Planck length scale. To estimate its parameters, note that when Hawking evaporation drives the black hole mass $M$ to the Planck scale, the corresponding Schwarzschild radius matches the Planck length,
\begin{equation}
R_{s} = \frac{2GM}{c^{2}} \sim \ell_{\rm Pl}, \quad M \sim M_{\rm Pl} = \sqrt{\frac{hc}{G}} \sim 10^{-5}~{\rm g}
\end{equation}
At this stage, quantum pressure balances further collapse and this leads to halting the evaporation and leaving behind a Planck star remnant. The key feature of PSR is that they behave externally as black holes but possess a non-singular LQG regulated interior.
\\
\\
The cosmological relevance of such objects becomes apparent when considering PBHs as these can form in the early universe when large-amplitude curvature perturbations re-enter the horizon during the radiation-dominated era. In the absence of new physics, PBHs lighter than $\sim 10^{15}~{\rm g}$ would have evaporated completely by now due to Hawking radiation. However, if evaporation halts at the Planck scale each PBH would instead leave behind a relic of mass $M_{\rm Pl}$. The present-day relic number density $n_{\rm relic}$ required to explain dark matter follows from
\begin{equation}
n_{\rm relic} = \frac{\rho_{\rm DM}}{M_{\rm Pl}} \sim \frac{2.3 \times 10^{-30}~{\rm g\,cm^{-3}}}{10^{-5}~{\rm g}} \sim 2.3 \times 10^{-25}~{\rm cm^{-3}}
\end{equation}
This corresponds to only a few hundred relics within the volume of the Earth, consistent with their being non-interacting and invisible to direct detection and yet sufficient to explain the gravitationally inferred dark matter density.
\\
\\
Another way to assess the role of PSR is in terms of entropy and temperature. A semiclassical black hole of mass $M$ evaporates at a characteristic Hawking temperature
\begin{equation}
T_{H} \sim \frac{1}{8\pi G M}
\end{equation}
As $M \to M_{\rm Pl}$, this formula becomes unreliable since quantum gravitational effects dominate and so in the PSR scenario, evaporation ceases before any runaway instability and the remnant enters a cold, non-radiating state. This stabilizing mechanism ensures that the PSR acts effectively as cold dark matter, redshifting like pressureless dust and contributing to the growth of large-scale structure without violating astrophysical bounds.
\\
\\
We note that the idea of Planck mass relics as dark matter candidates has a rich history, beginning with the proposal of MacGibbon \cite{mac1MacGibbon:1987my} and expanded upon in subsequent works \cite{mac2Maldacena:2020skw,mac3Rasanen:2018fom,mac4Barrau:2019cuo,mac5Chen:2004ft,mac6XENON:2023iku,ref1Arbey:2021mbl,ref2Easson:2002tg,ref3Pacheco:2018mvs,ref4Calza:2024fzo,ref5Calza:2024xdh,ref6Dialektopoulos:2025mfz}. More recently, related considerations have also appeared in \cite{Davies:2024ysj}. In these approaches, the motivation was largely semiclassical or thermodynamic with the invoking of a cutoff at the Planck scale to argue that evaporation halts, leaving stable relics. By contrast, the Planck star remnants we consider are not merely structureless endpoints but they emerge from a well-defined quantum gravitational process as the resolution of the classical singularity through a bounce in LQC. Their internal spacetime is described by an LQG-modified geometry, smoothly matched to a Schwarzschild exterior and their stability is guaranteed by quantum geometry effects rather than by external assumptions. This distinction places PSR in a fundamentally different category from the traditional MacGibbon-type relics, offering both a richer theoretical underpinning and potentially distinct phenomenology.
\\
\\
\section{Constraints from Collapse Fraction and Gravitational Waves}
\begin{figure} [!h]
    \centering    \includegraphics[width=1\linewidth]{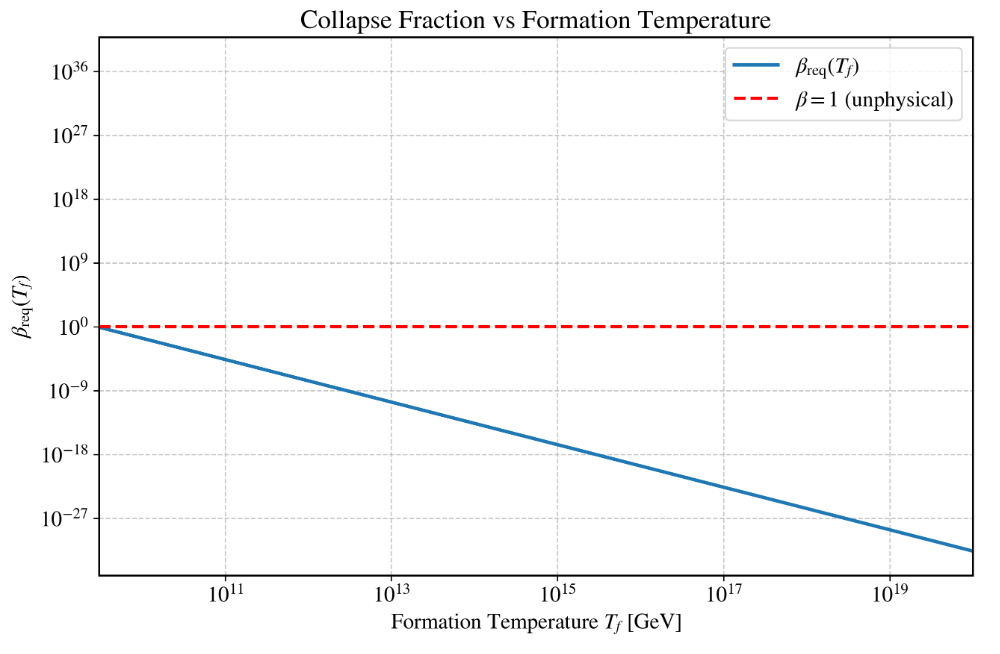}
    \caption{Required collapse fraction $\beta_{\text{req}}$ as a function of PBH formation temperature $T_f$, showing the viable window for PSR dark matter lies slightly above $10^9$ GeV.}
    \label{1}
\end{figure}

\begin{figure}[!h]
    \centering    \includegraphics[width=1\linewidth]{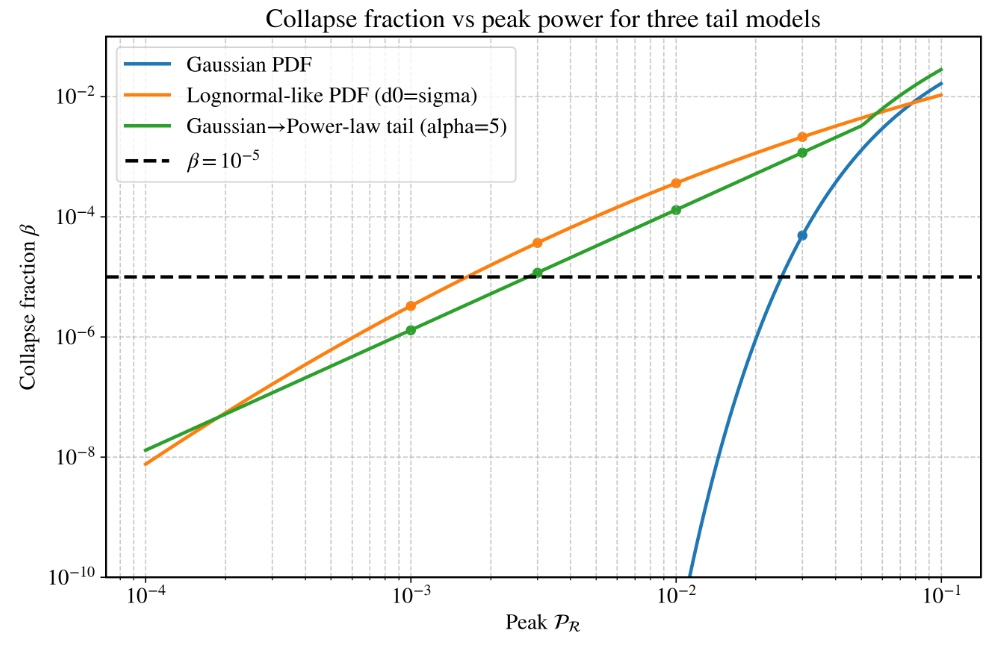}
    \caption{Collapse fraction $\beta$ as a function of peak curvature power $\mathcal{P}_\mathcal{R}^{\rm peak}$ for Gaussian, lognormal, and power-law-tailed statistics, with the target $\beta=10^{-5}$ indicated.}
    \label{2}
\end{figure}

To assess the viability of PSR as the dominant component of dark matter, we compute the required collapse fraction of primordial black holes PBHs during the radiation-dominated era and compare the resulting induced stochastic gravitational wave (GW) background with present observational limits. The starting point is the expression for the horizon mass at temperature $T$, given by
\begin{equation}
M_{H}(T) = 2 \times 10^{5} M_{\odot} \left(\frac{\mathrm{MeV}}{T}\right)^{2}
\end{equation}
If a fraction $\beta(T)$ of the radiation energy density at temperature $T$ collapses into PBHs that subsequently evaporate into Planck-scale relics of mass $M_{\rm PSR}$, the required fraction to explain the present dark matter density is
\begin{equation}
\beta_{\rm req}(T) = \frac{\rho_{\rm DM,0}}{\rho_{\rm rad}(T)} \, \frac{M_{H}(T)}{M_{\rm PSR}}
\left(\frac{T}{T_0}\right)^{3} \left(\frac{g_{*}(T)}{g_{*s,0}}\right),
\end{equation}
where $\rho_{\rm DM,0} \simeq 2.3 \times 10^{-47}\,\mathrm{GeV}^{4}$ is the present dark matter density, $\rho_{\rm rad}(T) = \frac{\pi^{2}}{30} g_{*}(T) T^{4}$ is the radiation energy density at temperature $T$, $T_{0} \simeq 2.35 \times 10^{-13}\,\mathrm{GeV}$ is the CMB temperature today, and $M_{\rm PSR} \simeq 2.18 \times 10^{-5}\,\mathrm{g}$ is the Planck mass. Evaluating this expression shows that $\beta_{\rm req}$ falls below unity around $T_{f} \sim 3 \times 10^{9}\,\mathrm{GeV}$ and reaches values $\beta_{\rm req} \sim 10^{-5}$ near $T_{f} \sim 10^{11}\,\mathrm{GeV}$. This defines the viable window for PSR dark matter, as illustrated in Figure \ref{1}. The corresponding formation times are $t \sim 10^{-21}$-$10^{-25}$ s, which is well before electroweak symmetry breaking and thermal leptogenesis and consistent with formation immediately after inflationary reheating. 
The collapse fraction is determined by the distribution of density perturbations at horizon re-entry and during radiation domination, the critical overdensity is
\begin{equation}
\delta_{c} \simeq 0.45
\end{equation}
as indicated by numerical simulations. For Gaussian statistics, the Press-Schechter expression yields
\begin{equation}
\beta(M_{H}) = \frac{1}{2} \, \mathrm{erfc}\!\left(\frac{\delta_{c}}{\sqrt{2}\,\sigma(M_{H})}\right),
\end{equation}
where $\sigma(M_{H})$ is the variance of the density contrast smoothed on the horizon scale $R_{H}$. The variance is related to the curvature power spectrum $P_{\mathcal{R}}(k)$ by
\begin{equation}
\sigma^{2}(R) = \frac{16}{81} \int_{0}^{\infty} d\ln k \, (kR)^{4} W^{2}(kR) \, P_{\mathcal{R}}(k),
\end{equation}
with $W(kR)$ a smoothing window function. For a narrow bump in the power spectrum, this reduces to the approximate relation
\begin{equation}
\sigma^{2}(R_{H}) \simeq \frac{4}{9} P_{\mathcal{R}}(k_{c}),
\end{equation}
where $k_{c}$ corresponds to the horizon scale at formation. Requiring $\beta \simeq 10^{-5}$ implies
\begin{equation}
\sigma(R_{H}) \simeq \frac{\delta_{c}}{\sqrt{2} \, \mathrm{erfc}^{-1}(2\beta)} \simeq 0.106,
\end{equation}
and therefore the required peak amplitude of the curvature power spectrum is
\begin{equation}
P_{\mathcal{R}}(k_{c}) \simeq 5.6 \times 10^{-2}.
\end{equation}
Thus, Gaussian statistics demand a localized enhancement in the curvature spectrum of order $10^{-2}$--$10^{-1}$ at comoving scales corresponding to $T_{f} \sim 10^{9}$--$10^{11}$ GeV.
\\
\\
Such large curvature perturbations inevitably generate a second-order stochastic GW background at horizon re-entry. The peak amplitude of the induced spectrum is approximately
\begin{equation}
\Omega_{\rm GW}^{\rm peak} h^{2} \simeq C(\sigma_{k}) \left(\frac{A}{10^{-2}}\right)^{2}
\end{equation}
where $A$ is the bump amplitude and $C(\sigma_{k}) \sim 10^{-8}$--$10^{-7}$ for narrow peaks. For $A \simeq 5.6 \times 10^{-2}$, one obtains
\begin{equation}
\Omega_{\rm GW}^{\rm peak} \sim 10^{-7}-10^{-6},
\end{equation}
peaking at frequencies $f \sim 100$--$1000$ Hz, and this is directly in the LIGO band. But it is imperative to note that this exceeds the LIGO O3 bound \cite{ligo}
\begin{equation}
\Omega_{\rm GW}(f) \lesssim 5 \times 10^{-9},
\end{equation}
by factors of $10$-$1000$, effectively ruling out the Gaussian scenario. The tension persists even at $T_{f} \sim 10^{11}$ GeV, unless $P_{\mathcal{R}}$ is suppressed to $\lesssim 10^{-3}$ and this is insufficient to yield the required collapse fraction.
\\
\\
To alleviate this conflict, one must consider departures from Gaussian statistics. For example, if the distribution of $\delta$ has heavy tails, the exponential suppression of the Gaussian case is avoided. Consider a lognormal-like distribution with shift parameter $d_{0}$ of order $\sigma$, in which case the collapse fraction scales as
\begin{equation}
\beta_{\rm lognormal} \sim \exp\!\left[- \frac{(\delta_{c} - d_{0})^{2}}{2\sigma^{2}}\right]
\end{equation}
This leads to a significantly larger $\beta$ for the same variance. For instance, at $P_{\mathcal{R}} \simeq 10^{-3}$ (so $\sigma^{2} \simeq 4 P_{\mathcal{R}}/9$) one finds $\beta_{\rm lognormal} \sim 3 \times 10^{-6}$, which is very close to the required value. Similarly, for a power-law tailed distribution
\begin{equation}
P(\delta) \propto \delta^{-\alpha}, \quad \delta > \delta_{c}
\end{equation}
with $\alpha \sim 5$, the collapse fraction becomes
\begin{equation}
\beta_{\rm power} \sim \int_{\delta_{c}}^{\infty} \delta^{-\alpha} d\delta \sim \delta_{c}^{-(\alpha-1)} \sim 10^{-6}
\end{equation}
which is again close to the required abundance for $P_{\mathcal{R}} \simeq 10^{-3}$. These results are summarized in Figure \ref{2}, which shows $\beta$ as a function of $P_{\mathcal{R}}$ for Gaussian, lognormal, and power-law cases. The contrast is stark as while Gaussian statistics require $P_{\mathcal{R}} \gtrsim 3 \times 10^{-2}$ to reach $\beta \sim 10^{-5}$, heavy-tailed non-Gaussian distributions achieve the same abundance with $P_{\mathcal{R}} \sim 10^{-3}$ and this also leads to reducing the induced GW signal below the LIGO bound.
The above analysis demonstrates that Gaussian primordial fluctuations in the radiation-dominated era cannot generate sufficient PBHs to explain dark matter as Planck mass relics without overproducing a stochastic GW background already excluded by LIGO. Scenarios with significant primordial non-Gaussianity naturally enhance the collapse fraction at fixed variance, relaxing the requirement on $P_{\mathcal{R}}$ and bringing the induced GW signal into compatibility with current limits and so if PSR constitute dark matter appreciably, their origin must involve strongly non-Gaussian fluctuations or an alternative non-standard cosmological history.
\\
\\
\section{Conclusions}
We have examined the viability of PSR as dark matter in light of the observational constraints imposed by the stochastic gravitational wave background. Beginning with the corrected collapse fraction formalism, we demonstrated that under Gaussian statistics the production of PBHs capable of leaving behind Planck relics at the required abundance necessarily demands a curvature power spectrum amplitude $P_{\mathcal{R}} \sim 10^{-2} - 10^{-1}$ at horizon re-entry. Such enhancements unavoidably generate a second-order gravitational wave background at frequencies in the LIGO band with $\Omega_{\rm GW} \sim 10^{-7} - 10^{-6}$, which exceeds current bounds by orders of magnitude.
\\
\\
The tension can be alleviated if the primordial perturbations exhibit significant non-Gaussianity as heavy-tailed distributions, such as lognormal or power-law forms, enhance the collapse fraction $\beta$ at fixed variance, thereby reducing the required $P_{\mathcal{R}}$ and suppressing the corresponding induced gravitational wave background. In this case, PSR remain possible dark matter candidates and their existence would imply that the origin of small-scale fluctuations is inherently non-Gaussian. If dark matter is ultimately found to be comprised of Planck scale relics, it will point directly to non-trivial early universe physics beyond standard slow-roll inflation and Gaussian statistics.
\\
\\
\section*{Acknowledgements}
OT was supported in part by the Vanderbilt Discovery Alliance Fellowship. AL was supported in part by the Black Hole Initiative, which is funded by GBMF and JTF.
\bibliography{references}
\bibliographystyle{unsrt}

\end{document}